\documentstyle[preprint,aps]{revtex}

\input epsf

\begin{document} 

\draft 

\preprint{}

\def\r{{\bf r}}   \def\q{{\bf q}}

\title{Structure Factor of a Lamellar Smectic Phase with Inclusions}

\author{Pierre Sens\dag\ \& Matthew S. Turner\ddag\S} 

\address{\dag Institut Charles Sadron - 6 rue
Boussingault, 67083 Strasbourg Cedex, France\footnote{email: sens@ics.u-strasbg.fr}\\  
\ddag Rockefeller University, New York, NY 10021 USA and\\ 
\S Department of Physics, Warwick University, Coventry, CV4
7AL, UK} 

\maketitle

\begin{abstract}

Motivated by numerous X-ray scattering studies of lamellar phases with membrane proteins, amphiphilic peptides, polymers, or other inclusions, we have determined the modifications of the classical Caill\'e law for a smectic phase as a function of the nature and concentration of inclusions added to it. Besides a fundamental interest on the behavior of fluctuating systems with inclusions, a precise characterization
of the action of a given protein on a lipid membrane (anchoring, swelling, stiffening ...) is of direct biological interest and could be probed by way of X-ray measurements. As a first step we consider three different couplings involving local pinching (or swelling), stiffening or tilt of the membrane. In the first two cases we predict that independent inclusions induce a simple renormalization of the bending and
compression modulii of the smectic phase. The X-ray experiments may also be used to probe correlations between inclusions. Finally we show that asymmetric coupling (such as a local tilt of the membrane) results in a modification of the usual Caill\'e law.

\end{abstract} 

%

%
\pacs{87.22.Bt:
Membrane and subcellular physics and structure\\ 82.65.Dp: Thermodynamics of surfaces and interfaces\\
82.70.y: Disperse systems\\  submitted to {\em Euro. Phys. J E} on 09/03/2000} 
%
%

\narrowtext

Lyotropic smectic phases ($L_\alpha$ phases) are liquid crystalline systems of 
parallel, regularly spaced lipid bilayers\cite{mmmm,condmat}. Those phases form a quasi-crystalline structure in one (the $z$) direction, while retaining their fluid properties in the two others directions ($x-y$ plane). They have been studied extensively in recent years, as they form one of the most convenient lyotropic structure for the experimental study of fluctuating interfaces. They have been for instance particularly useful in demonstrating the importance of the so-called entropic (Helfrich) repulsion between fluctuating membrane\cite{safiroux}. Furthermore, the degree of dilution (the layer spacing) and other controlled parameters of the lyotropic $L_\alpha$ phase can be experimentally varied over a wide range, which allows for a precise study of the scattering intensity of the phase. This scattering intensity is expected to show several peaks corresponding to a
 stack of regularly spaced membrane. In these layered systems however, the thermal fluctuations destroy the long range order at finite temperature, leading to weak - power law - singularities of the peaks. 

Inclusions in membrane have been actively studied for many years. Most of the earlier work was
done by the biological community, for which it was a necessary step toward the comprehension of the
complexity of biological (cell) membrane\cite{alberts}. In the last twenty years or so, a more physical description of membrane-inclusion complexes, in term of curvature energy and disruption of the molecular order of the membrane, have led to interesting results on membrane-induced interactions between inclusions\cite{goulian}. While this ``physical'' picture of the membrane-inclusion interactions leaves aside many fundamental processes driven by specific interactions, it allows for a description of universal features of the complexes in terms of elastic couplings between membrane and inclusion. This description is applicable to a certain extent to membrane with adsorbed or grafted polymers or colloids\cite{lipo,jf}, as well as to transmembrane inclusions resembling certain membrane proteins\cite{goulian}.

In recent years, numerous experimental studies have investigated the role of 
added membrane proteins in lyotropic smectic phase. This lipid phase seems to 
be of particular interest for the study ``in vitro" 
of the bilayer-protein association\cite{safi1,urbach,safi2,taulier,tsapis},
 and may have some practical application for gene therapy\cite{safi3}. 
 From a physicist point of view, it is interesting to understand the 
 effect of inclusions on the thermal fluctuations of flexible, fluid membranes. 
 From a more ``biological'' point of view, a precise characterization of the physical 
 action of a given protein onto a lipid membrane (anchoring, swelling, stiffening ...) 
 is of direct biological interest, and could be reach via X-ray measurements. This work
  have some relevance in material science as well, since for practical purposes, a lamellar phase is very rarely composed of two components (solvent+surfactants) only. 
The adjunction of a third or fourth component, such as cosurfactant, polymers, proteins or colloids is of common practice to tune the properties of the phase\cite{lalphaplus,ligoure}. The object of the present work is to determined the modifications of the smectic X-ray scattering due to the presence of inclusions in the lamellar phase. Ultimately,  one could expect 
to map out different membrane-inclusion couplings by their effect on the X-ray structure factor. We consider three model inclusions, leading to different modifications of the membrane
 elastic properties, and we show that the qualitative analysis of the diffraction 
 spectrum is enough to discriminate between these inclusions.
 
Several recent theoretical works deal with lamellar systems with flexible polymers\cite{ligoure} or other soluble inclusions 
(so-called doped solvent lamellar phases). These works are complementary, as we are 
interested in insoluble inclusions such as the membrane proteins used in \cite{urbach}, 
while \cite{ligoure} studied effects such as polymer depletion - induced interactions
 between lamellae, resulting from an equilibrium between the trapped inclusions and free inclusions in pure solvent.

\vskip0.7truecm$\diamondsuit$ The standard description of Caill\'e\cite{caille} for pure $L_\alpha$ phases has been confirmed experimentally in numerous systems. It is based on the classical smectic
hamiltonian\cite{degenne}: \begin{equation} H_0={1\over 2}\int d^3r\left(B\left(\partial_zu\right)^2+K\left(\nabla_\bot^2
u\right)^2\right) \label{hamilt0}
\end{equation} where $u(\r)$ is the (continuous) normal displacement of the layer at the point $\r$, $z$ is the coordinate normal to the layers, and $\nabla_\bot$ is the gradient along the layers. The elasticity of the smectic phase is characterized by the two coefficients $K$ and $B$, which are related
to the energy cost of bending and compressing the sample. Typically, $K=\kappa/d$, where $\kappa$ is the bending modulus of a single membrane and $d$ is the average layer spacing, and $B$ is function of the intermembrane interactions\cite{mmmm}. Those modulii define the characteristic smectic length $\lambda=\sqrt{K/B}$\cite{degenne}. The scattering intensity $I(\q)$ is related to the Fourier transform
of the density correlation function $G_n(\r)$, itself related to the exponential of the layer displacement correlation function $g(\r)=q_0^2/2\langle\left(u(\r)-u(0)\right)^2\rangle$, where
$\langle...\rangle$ represents a thermal average of the fluctuations, and $q_0=2\pi/d$ (see \cite{condmat}, sec 6.3.2). As usual, the Fourier decomposition of the layer displacement is used: $u({\bf r})=\int d^3q/(2\pi)^3 u_q e^{i{\bf q r}}$. Using Eq.(\ref{hamilt0}): 
\begin{equation} g(\r)={q_0^2\over V}
\int{d^3q\over(2\pi)^3}\langle|u_q^2|\rangle\left(1-e^{i\q \r}\right)\quad{\rm with}\quad
\langle|u_q^2|\rangle_0={V k_BT\over B q_z^2+K q_\bot^4} \label{rms0} 
\end{equation} 
where $V$ is the volume of the sample. After integration
over $q_z$, an asymptotic solution of $g(\r)$ can be found\cite{niels}: 
\begin{eqnarray}
g^{(0)}(\r)={q_o^2 k_BT\over 4\pi\sqrt{KB}} \int dq_\bot{\left(1-J_0(q_\bot r_\bot)e^{\textstyle
-\lambda z q_\bot^2}\right)\over q_\bot}\simeq\eta\left(2\gamma+2\log{\pi r_\bot\over
l}+E_1\left({r_\bot^2\over 4\lambda z}\right)\right)&\cr 
{\rm
with}\qquad\eta\equiv{q_0^2k_BT\over8\pi\sqrt{KB}}\hskip6truecm& 
\label{rms02} 
\end{eqnarray} 
where $J_0(x)$ is the Bessel function of the first kind, $\gamma$ is the Euler constant, and $E_1(x)$ is the
exponential integral ($l$ is a molecular size). From Eq.(\ref{rms02}), one can show that the scattering
intensity $I_n(\q)$ near the n$^{\rm th}$ peak, defined by 
\begin{equation} 
I_n(\q)=G_n(\q+n\q_0)+G_n(\q-n\q_0)\quad{\rm with}\quad
G_n(\r)\propto e^{-n^2 g(\r)} 
\label{intensity} 
\end{equation}  
shows power law singularities of the
form\cite{caille,gunther} 
\begin{equation} 
G_n(\q-n\q_0)\sim\cases{(q_z-n q_0)^{-2+\eta} \hskip
0.7truecm{\rm if}\quad q_\bot=0\cr q_\bot^{-4+2\eta} \hskip 2.2truecm{\rm if}\quad q_z=0} 
\label{peaks}
\end{equation} 

The above law gives a satisfactory description of the X-ray structure factor in the
vicinity of the diffraction peaks, and has been observed experimentally (see \cite{mmmm} and references
therein). A more accurate calculation of $I(\q)$ over a broad range of wave vector requires a discrete
description of the lamellar phase\cite{bruinsma}, and the account of the fluctuations of the lipid concentration\cite{nallet}, and is not tractable analytically. As one can see, this peculiar power law
divergence allows for an experimental determination of the product $KB$, through the parameter $\eta$. Low values of the parameter $\eta$, such as those observed for electrostatically stabilized systems ($\sim 0.2$) correspond to a well defined smectic organization with little fluctuations. Larger values of order unity are observed for weakly interacting neutral or screened systems stabilized by the Helfrich interaction\cite{mmmm}.

\vskip0.7truecm$\diamondsuit$ Several couplings between the inclusions and the membrane are considered below. They are referred to as ``pinch", ``stiff", and ``tilt", and are schematically depicted in Fig.1 and Fig.2. The ``pinch" inclusion can be thought of 
as exerting a force which tends to pinch ($\beta_{pinch}>0$, see Eq.(\ref{hamiltpinch}) below) or swell ($\beta_{pinch}<0$) neighboring membranes. Since we are using a continuous description for the lamellar phase, the inclusion may reside within a membrane or between two membranes. Inclusions which would typically induce such deformations are amphiphilic proteins\cite{urbach}, with hydrophobic parts which penetrate in the bilayers and hydrophilic parts lying in the water, or simply particles larger than the layer spacing which sterically swell the membrane. A pinched lamellar structure has also been identified in the case of anionic polymers added to a lamellar phase formed by a mixture of cationic and neutral lipids\cite{subu}. The hamiltonian corresponding to this perturbation couples the density of inclusion $\rho(\r)$ with the variation of the layer spacing\cite{us}: 
\begin{equation} 
\Delta
H_{pinch}=\int d^3r\beta_{pinch}\rho(\r)\partial_zu 
\label{hamiltpinch} 
\end{equation} 

The ``stiff" inclusion corresponds to a transmembrane protein which locally stiffens (or softens) the membrane and modifies its fluctuations. A stiffening is expected for many integral membrane proteins, or for rigid inclusions laying on the lipid bilayer. On the other hand, pores in the membrane are expected to induce a softening of the membrane (among other, less trivial phenomena). The inclusion density is coupled with a variation ($\delta K$) of the bending modulus\cite{them,netz}: 
\begin{equation} 
\Delta H_{stiff}={1\over2}\int d^3r\delta K v\rho(\r)\left(\nabla_\bot^2u\right)^2 
\label{hamiltstiff} \end{equation} 
where $v$ is the effective volume of an inclusion in the membrane.

The ``tilt" inclusion corresponds to an anisotropic inclusion (such as a grafted polymer) which would locally induce a curvature to a membrane without changing the layer spacing. It has been shown that such a deformation can be described by a modification of the ``pinch" hamiltonian Eq.(\ref{hamiltpinch})\cite{us2}; we will get
back to it later. It is clear that any real inclusion will, to a certain extent, combine the three
effects described above.

A single inclusion of the ``pinch'' or ``tilt'' type induces a non-zero average deformation to the membrane. However, this mean deformation evens out during the course of an experiment, and is not detectable on the scattering spectrum (except if it induces a phase separation, such as expulsion of the solvent\cite{taulier,subu}). The effect of the inclusions on the spectrum is visible via their effect on the membrane fluctuations. In what follows, we do not investigate a possible coupling between the membrane fluctuations and the inclusion correlation function (assumption of a quenched distribution of inclusions). It has been shown for a single membrane system that ``quenched'' and ``annealed'' distributions of inclusions give similar corrections to the fluctuation spectrum of the membrane\cite{netz}. The scattering experiments show an average of the structure factor over the positions of the inclusions, it is thus necessary to calculate
the membrane correlation function after a proper average over the possible positions of the inclusions.  

\vskip0.7truecm$\diamondsuit$ For the case of ``pinch"
inclusions, the hamiltonian Eqs.(\ref{hamilt0},\ref{hamiltpinch}) can be rewritten in the discrete form:
\begin{eqnarray} H_{pinch}={1\over V}\sum_{q>0}\left\{Q_q\left|u_q u^*_q\right|^2-\left(\xi_q
u_q^*+\xi^*_q u_q\right)\right\}\cr{\rm with}\quad Q_q=B q_z^2+K q_\bot^4\quad{\rm and}\quad \xi_q=\beta_{pinch}
i q_z \rho_q \label{hqpinch} \end{eqnarray}

The structure factor is then easily obtained: \begin{equation} {\langle|u_q|^2\rangle\over V}={T\over
Q_q}+{\xi_q\xi_q^*/V\over Q_q^2}={k_BT\over Bq_z^2+Kq_\bot^4}
+\beta_{pinch}^2{q_z^2\over\left(Bq_z^2+Kq_\bot^4\right)^2}{\langle|\rho_q|^2\rangle_\rho\over V}
\label{uqpinch} \end{equation} where $\langle...\rangle_\rho$ designates an average over the position of
the inclusions. The first part of the left hand side corresponds to the correlation function in a pure
$L_\alpha$ phase $g^{(0)}(\r)$ (Eq.(\ref{rms0})), and the second part describes the corrections due to
the inclusions $\Delta g$. This correction reflects correlations between inclusions
$\langle|\rho_q|^2\rangle_\rho$.

If the inclusions are independent from each other, $\langle|\rho_q|^2\rangle/V$ is the average
concentration of inclusions $\bar\rho$. Similarly to Eq.(\ref{rms02}), the correction to the spatial
structure factor can be written: \begin{equation} \Delta g(\r)=2\eta{\beta_{pinch}^2\bar\rho\over 2 k_BT
B}\left(\int dq_\bot{\left(1-J_0(q_\bot r_\bot)e^{\textstyle -\lambda z q_\bot^2}\right)\over
q_\bot}+{1\over 4}e^{-{{\textstyle  r_\bot^2}\over{\textstyle 4\lambda z}}}\right) \label{dgpinch}
\end{equation}

This correction leads to the same power law divergence of the scattering intensity as the pure
$L_\alpha$ (Eq(\ref{peaks})), with the effective modulii: \begin{equation}
\left(KB\right)_{pinch}={\textstyle KB\over \left(1+{\beta_{pinch}^2\bar\rho\over 2k_BT B}\right)^2}
\label{renorm0} \end{equation}

One notice right away that the ``pinch" inclusions tend to soften the lamellar phase. We thus expect
broader peaks in the scattering intensity (see Fig.1). This finding is qualitatively consistent with experiments on
comparable systems\cite{urbach,urbach2}. Note that results reported for the doped solvent situation tend
to show qualitatively similar results for physically different reasons. In that case, the inclusions-mediated
attraction between lamellae tends to reduce the compression modulus $B$\cite{ligoure}.

Possible correlations between inclusions may be induced either by direct interactions between them, or by membrane mediated interactions. In the Latter case, it has been shown that the inclusions interact via
the potential (in Fourier space)\cite{us} 
\begin{equation} U_q=-{\beta_{pinch}^2\over 2}{q_z^2\over Q_q}
\label{pot} 
\end{equation} 
In this case, a "Debye-Huckel like"\cite{condmat} expansion of the correlation between
inclusions (valid if $\bar\rho U_q/(k_BT)\ll1$), shows that: 
\begin{eqnarray} 
{\langle|\rho_q|^2\rangle\over
V}\simeq{\bar\rho\over 1+\bar\rho U_q/(k_BT)}\quad{\rm or}\quad {\langle|\rho_q|^2\rangle\over
V}={\bar\rho Q_q\over Q_q^\prime}\cr {\rm with}\quad Q_{q^\prime}\equiv B'q_z^2+Kq_\bot^4\quad{\rm
and}\quad B'\equiv B\left(1-{\bar\rho\beta_{pinch}^2\over 2 k_BT B}\right) 
\label{cor2} 
\end{eqnarray}

From Eq.(\ref{uqpinch}), the corrections to the structure factor is given by: \begin{equation} \Delta
\langle|u_q|^2\rangle={\beta_{pinch}^2 q_z^2\bar\rho\over
Q_qQ_q^\prime}\quad\rightarrow\quad\eta_{pinch}=\eta\left(1+{\beta_{pinch}^2\bar\rho\over
k_BT\sqrt{B'}(\sqrt{B}+\sqrt{B'})}\right) \end{equation}

The effective modulii are now given by: 
\begin{equation}
\left(KB\right)_{pinch}={KB\over \left(1+\frac{\beta_{pinch}^2\bar\rho}{2k_B TB}+{3\over 16}\left({\textstyle
{\beta_{pinch}^2\bar\rho\over 2k_BT B}}\right)^2\right)} 
\label{renorm1} 
\end{equation}
Note that in the limit of validity of the expansion, the correction due to correlation is quite small.

\vskip0.7truecm$\diamondsuit$ The investigation of the effect of the ``stiff" inclusions is more delicate, because of the non-quadratic nature of the hamiltonian Eq.(\ref{hamiltstiff}). This leads to a coupling between
different Fourier modes of the membrane displacement, in which case Eq.(\ref{rms0}) for the real space
correlation function has to be replaced by: 
\begin{equation} 
g(\r)={1\over 2}q_0^2 \int{d^3q\over(2\pi)^3}{d^3q^\prime\over(2\pi)^3}\langle
u_qu^*_{q^\prime}\rangle\left(e^{i\q \r}-1\right)\left(e^{-i\q^\prime \r}-1\right) \label{rms03}
\end{equation} The discrete version of the hamiltonian reads: 
\begin{eqnarray} 
{\textstyle{H_{stiff}}\over
k_BT}=&\sum_{q,q^\prime>0}\left(G^{0}_{q,q^\prime}+\Delta G_{q,q^\prime}\right)u_q
u^*_{q^\prime}\hskip3truecm&\cr
=&{1\over2}\sum_{q,q^\prime}\left({\textstyle{Q_q}\over \textstyle{V k_BT}}\delta_{q,q^\prime}+{\textstyle{\delta K v}\over \textstyle{k_B T
V^2}}q_\bot^2 q^\prime_\bot{}^2\rho_{-(q-q^\prime)}\right) u_q u^*_{q^\prime}&
\label{discretestiff}
\end{eqnarray} 
from which the mean square displacement can be calculated: \begin{equation} \langle
u_qu^*_{q^\prime}\rangle=\left[\left(G^{0}+\Delta G\right)^{-1}\right]_{q,q^\prime} \label{matrix}
\end{equation} 
In the limit where the correction is only a small perturbation, an expansion of the
previous equation $\left[G^{0}+\Delta G\right]^{-1}_{q,q^\prime}=
\left[G^{0}\right)]^{-1}_{q,q^\prime}-\left[\Delta G\right]_{q,q^\prime}\times
\left[G^{0}\right]^{-2}_{q,q^\prime}$ leads to: 
\begin{equation} 
{\langle u_qu_{q^\prime}^*\rangle\over V}\simeq{k_BT \over B q_z^2+K q_\bot^4}\delta_{q,q^\prime}-{\delta K v k_BT\over V} {q_\bot^2q^\prime{}_\bot^2\rho_{q-q^\prime}\over\left(B q_z^2+K q_\bot^4\right)\left(B q^\prime_z{}^2+K
q^\prime_\bot{}^4\right)} \label{uqstiff} \end{equation}

For an homogeneous distribution of particle of average concentration
$\bar\rho$, the Fourier modes are decoupled: $\rho_q=\bar\rho\delta_q$. It is then easy to show that
again, the scattering intensity show peaks with power law divergence, from which the product of the
effective compression and bending modulii can be extracted. The renormalization introduced by the
inclusions is given by: \begin{equation} \left(KB\right)_{stiff}={KB\over \left(1-{\textstyle{\delta K
v\bar\rho}\over \textstyle{2K}}\right)^2} \label{renormstiff} \end{equation} Note that this results
compares to the finding of Netz and Pincus\cite{netz}, who have shown that for a single membrane with inclusions, the effective bending
rigidity of the membrane is given by $\kappa_{eff}\sim\kappa+\delta\kappa\phi$ at the lowest order in the volume fraction $\phi$ of inclusions.

In contrary to the case of ``pinch" inclusions, a stiffening of the $L_\alpha$ phase is observed for
inclusions increasing locally the bending rigidity of the membrane. This should lead to a sharpening of the peaks in the scattering intensity. Hence, a quick look at the intensity scattered by the
sample should give us valuable information on the inclusion-membrane coupling. A stiffening of a
L$_\alpha$ phase has been recently observed upon adjunction of polysoap molecules\cite{carlos}. This is
a more complex case, since the polysoaps are fairly long molecules and can be expected to modify
(increase) the interactions between membranes in addition to their effect on an isolated bilayers. The
reduction of the effective $\eta$ has however been observed below the overlap concentration of
polysoaps, where the molecules are expected to be spread flat onto the membrane.

\vskip0.7truecm$\diamondsuit$ The last coupling investigated in this work is the case of anisotropic inclusions which induce a local
tilt (spontaneous curvature) on a membrane. It has been shown in a previous work\cite{us2} that such a
tilt can be conveniently mimic with the aid of an analogy between the ``pinch/swell'' inclusions and +/- electrical charges in electrostatics (see Fig.2). The deformation of the smectic induced by a ``dipole'' $p_z\equiv \beta_{tilt} d$ (where $\beta_{tilt}$, the tilt intensity, has the dimension of an energy) normal to the membrane has been calculated\cite{us2}: 
\begin{equation} 
u_{tilt}={p_z\over 16\pi
\sqrt{KB}z^2}\left(1-{r_\bot^2\over4\lambda z}\right) e^{\textstyle -r_\bot^2/(4\lambda z)} \label{udip}
\end{equation} 
and correspond to imposing a local curvature $C_0=\beta_{tilt}/(\pi\kappa d)$ at the
position of the inclusion ($r_\bot=0$, $z=d/2$) without changing the layer spacing ($\kappa=Kd$ is the
rigidity of an isolated membrane). It can be shown easily in the Fourier space that such a deformation
can be induced by a coupling between the second order z-derivative of the layer deformation and the
inclusion density: 
\begin{equation} 
\Delta H_{tilt}=\int d^3r p_z\rho(\r)\partial^2_zu
\label{hamilttilt} 
\end{equation} 
which, for independent inclusions, leads to the membrane structure factor:: 
\begin{equation} 
{\langle|u_q|^2\rangle\over V}={k_BT\over Bq_z^2+Kq_\bot^4}
+p_z^2{q_z^4\over\left(Bq_z^2+Kq_\bot^4\right)^2}\bar\rho 
\label{uqtilt} \end{equation} and the
correlation function: 
\begin{equation} 
\Delta g(\r)={q_o^2 p_z^2\bar\rho\over 2\pi
B^2}\int{d^2q_\bot\over 2\pi}\int{dq_z\over2\pi}\left(1-e^{-i\q\r}\right){q_z^4\over (q_z^2+\lambda^2
q_\bot^4)^2} 
\end{equation} 
After integration over $q_z$ (the ($\bot$) indexes are
forgotten): 
\begin{eqnarray} 
\Delta g(\r)=-{q_o^2p_z^2\lambda\bar\rho\over8\pi B^2}{\cal
I}\hskip4.6truecm\cr {\rm where}\quad{\cal I}=3\int dq q^3\left(1-J_0(q r)e^{-\lambda z
q^2}\right)+\lambda z\int dq q^5 J_0(q r) e^{-\lambda z q^2} 
\end{eqnarray} 
Both integrals can be computed after a series expansion of the Bessel function. The correction to the correlation function
takes the form: 
\begin{equation} 
\Delta g(\r)={q_o^2p_z^2\lambda\bar\rho\over16\pi
B^2}\left({e^{-\omega}\over(\lambda z)^2}\left(1+\omega-\omega^2\right)+{\rm constant}\right)\quad{\rm
with}\quad\omega\equiv{r_\bot^2\over4\lambda z} 
\end{equation}

The X-ray spectrum is obtained by a combination of the above expression with
Eqs.(\ref{rms02},\ref{intensity}): 
\begin{eqnarray} 
G_\q\propto\int d^3r e^{i\q\r} \left({l\over
r_\bot}\right)^{2\eta} \exp{\left(-\eta\left(E_1(\omega)+\left({l_{tilt}\over
z}\right)^2e^{-\omega}\left(1+\omega-\omega^2\right)\right)\right)}\cr
{\rm with}\quad l_{tilt}\equiv
d\sqrt{\beta_{tilt}^2\bar\rho\over2k_BT B}\quad{\rm
and}\quad\omega\equiv{r_\bot^2\over4\lambda z} \hskip2.5truecm 
\label{bigint} 
\end{eqnarray}

Unlike the case of symmetrical inclusions considered earlier,  the asymptotic scaling laws for $G_\q$ in the limits $q_z=0$ or $q_\bot=0$ cannot be easily extracted from the above expression, since the exponential term in the integral cannot be written as a function of $r_\bot^2/(\lambda z)$ alone. However, up to the first order in the dipole strength, $l_{tilt}/z<1$ for $z>d/2$ and we can expand the exponential in the integral Eq.(\ref{bigint}). One can see clearly that unlike symmetrical
inclusions such as ``pinch'' and ``stiff'', the effect of a ``tilt'' inclusion cannot be matched by a renormalization of the smectic modulii, as they modify of the power law divergence of the scattering intensity near the Bragg positions. In fact, the inclusions add a term to the intensity,
which for $q_\bot=0$ can be written: 
\begin{eqnarray} 
I_{tilt}(q_z)\propto -q_z^\eta\
l_{tilt}^2l^{2\eta}\lambda^{1-\eta} \int{dZ\over
Z^{1+\eta}}e^{iZ}\int{d\omega\over\omega^\eta}e^{-\omega} \left(1+\omega-\omega^2\right) e^{-\eta
E_1(\omega)}\quad;\quad Z\equiv q_z z 
\end{eqnarray} 
and in the other limit $q_z=0$:
\begin{eqnarray} 
I_{tilt}(q_\bot)\propto -q_\bot^{2\eta}\ l_{tilt}^2l^{2\eta}\lambda\int{dR\over
R^{1+2\eta}}J_0(R)\int d\omega e^{-\omega} \left(1+\omega-\omega^2\right) e^{-\eta E_1(\omega)}\quad;\quad R\equiv q_\bot r_\bot 
\end{eqnarray}

The integrals being prefactors only, the scattering intensity should take the form:

\begin{eqnarray} 
I_{tilt}\propto\cases{ \alpha_1 q_z^{-2+\eta}-\alpha_2 q_z^\eta &for $q_\bot=0$\cr 
\alpha_1
q_\bot^{-4+2\eta}-\alpha_3 q_\bot^{2\eta} & for $q_z=0$\cr} \label{iqtilt} 
\end{eqnarray} 
where the $\alpha_i$ are constants. The first term is the Caill\'e law for a pure smectic phase and the second term is the signature of the asymmetric inclusions. Note that the effect of the tilt inclusion as described by
Eq.(\ref{iqtilt}) is most probably very small and might be hardly detectable in experiments. Note also
that as a ``pinch'' inclusion lead to a renormalization of the compression modulus $B$ (see Eq.(\ref{cor2})), a ``tilt'' inclusion can be seen as renormalizing the elastic modulus of an higher order term of the smectic hamiltonian, of the form $K^\prime(\partial_z^2u)^2$, which has been neglected in the reference
hamiltonian Eq.(\ref{hamilt0})\cite{lub}. This reinforces the fact that their effect on the structure
factor must be small. One nevertheless might hope to detect such an effect because of its dependence on the concentration of inclusion.

\vskip0.7truecm$\diamondsuit$ In conclusion, we have derived the effect of inclusions on the shape of the scattering curve of a smectic $L_\alpha$ phase. We have considered the three inclusions depicted in Figs.1 and 2, and have shown that the two symmetrical inclusions (``pinch'' and ``stiff'') renormalize the elastic modulii of the phase without modifying the power law divergence near the peaks (Caill\' e law). In agreement with experiments, we have obtained that the ``pinch'' inclusions soften the lamellar phase\cite{taulier}, while the ``stiff'' inclusions makes it more rigid\cite{tsapis}. the ``tilt'' inclusions are predicted to modify the shape of the scattering curves, but their effect is small and might hardly be detectable.

As a last remark, we note that in view of our results, the study of the 
X-ray scattering spectrum of a lamellar phase with inclusions should 
be a powerful tool to investigate some changes in conformation of membrane 
proteins. One can argue that the bridge (pinch) between neighboring membranes 
created by some amphiphilic peptides and reported in \cite{taulier} may 
be pulled out of one of the two membranes as the intermembrane spacing is 
increased. If the peptide is flexible enough, it is then expected to lay
 flat on one membrane\cite{tsapis}. This conformation change represents a 
 transition from a ``pinch'' to a ``stiff'' inclusion, and should be directly
  detectable from the shape of the scattering peak. Moreover, some amphiphilic 
  helical peptides in contact with a lipid bilayer are known to go from an 
  adsorbed state at low peptide concentration, to an inserted state (a certain 
  number of peptide aggregating to form an hydrophilic pore) at higher concentration\cite{huang}. This transition from one state to the other could 
be interpreted in terms of ``stiff" and ``soft'' (in the case of a pore) 
inclusions; One could investigate X-ray scattering from such a solution with 
this transition in mind. Nevertheless, one should always bear in mind that the different states of the amphiphilic peptides are most probably a complex mixed 
between the types of inclusion proposed in our work, and probably others.

\vskip1truecm
We acknowledge stimulating discussions with W. Urbach, N. Taulier and N. Tsapis (E.N.S Paris), and we would like to thank J.-F. Joanny and A. Johner (I.C.S Strasbourg) for fruitful comments.

\newpage

\vskip2truecm

\centerline{ \epsfxsize=15truecm \epsfbox{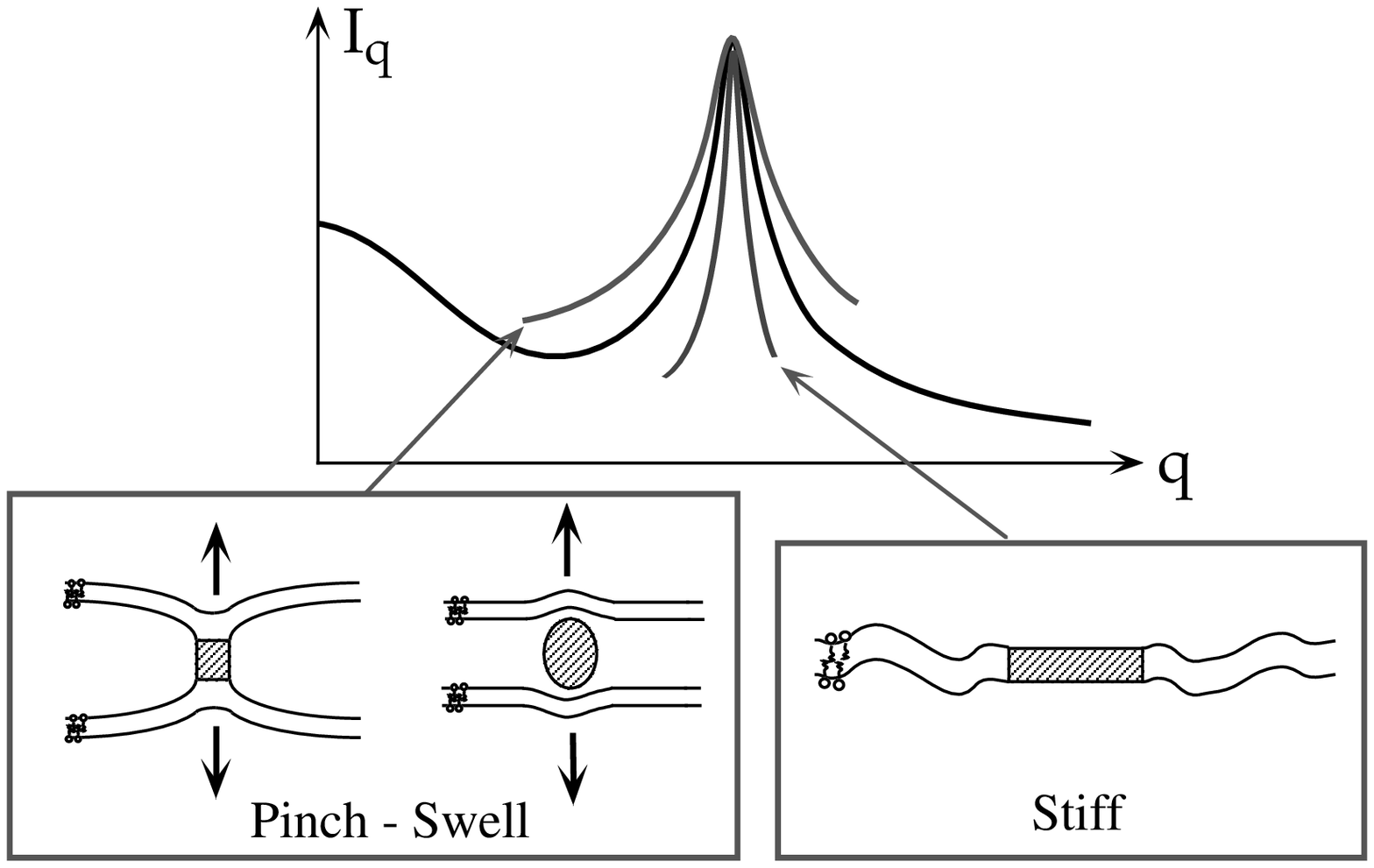} } 
\vskip-4truecm
\centerline{Fig.1 The two symmetrical inclusions considered in the paper. A ``pinch'' inclusion modifies}
\centerline{the layer spacing and broadens the Bragg peak - A ``stiff'' inclusion increase the local rigidity}
\centerline {and makes the Bragg peak narrower} 
\vskip1truecm 

\pagebreak

\centerline{ \epsfxsize=15truecm \epsfbox{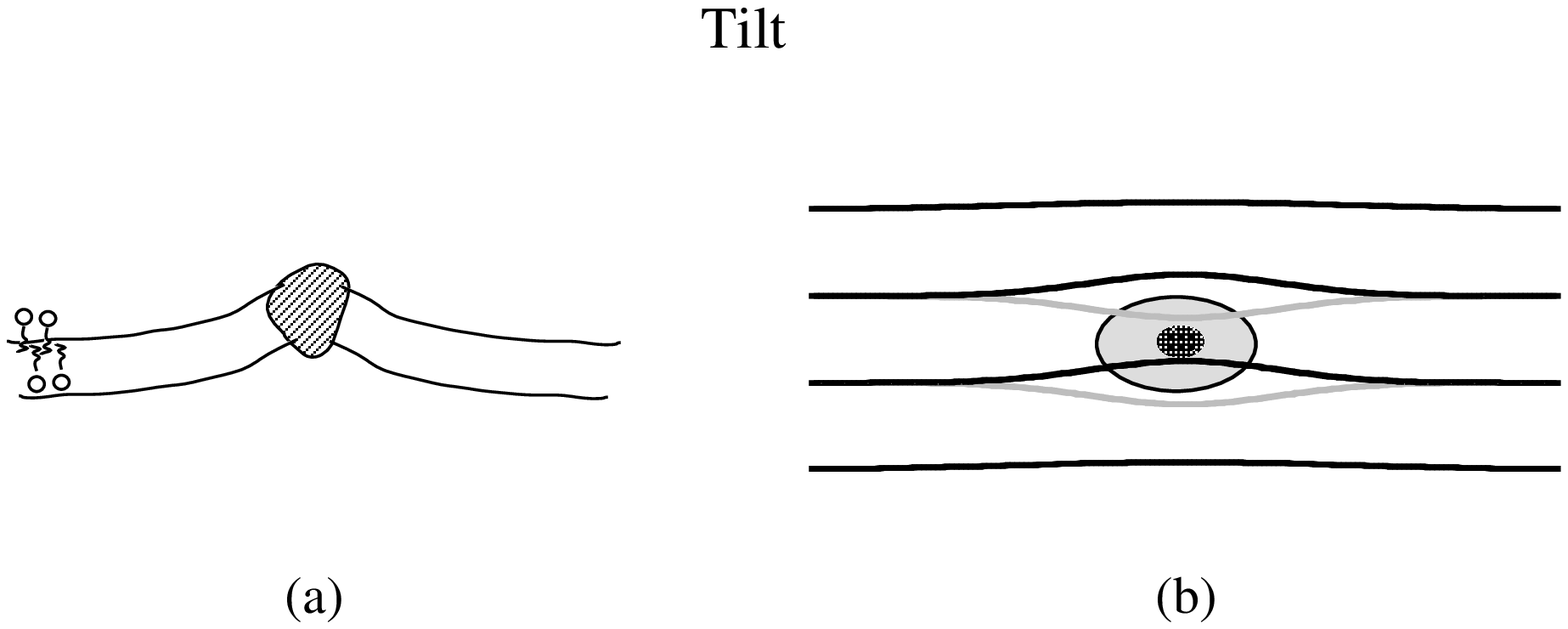} } 
\vskip-5truecm
\centerline{Fig.2 a) The ``tilt'' inclusion induces a spontaneous curvature}
\centerline{b) How to make a ``tilt'' from the combination of ``pinch'' and ``swell''\cite{us2}}


\begin{thebibliography}{99}
\bibitem{lub} The authors would like to thank T. Lubensky for this comment
\end{thebibliography}
\end{document}